\documentclass[12pt]{article}
\usepackage{amsfonts}
\usepackage{latexsym}
\usepackage{amsmath,amssymb}
\usepackage{verbatim}
\usepackage{setspace}
\usepackage{cite}
\usepackage[textheight=9in, textwidth=6.5in, letterpaper]{geometry}
\usepackage{graphicx}
\usepackage{dcolumn}
\usepackage{bm}
\usepackage{epsfig,amsmath}
\usepackage{float}
\usepackage[usenames,dvipsnames]{color}
\usepackage[linktocpage,pagebackref]{hyperref}

\begin{document}
\newcommand {\nn}    {\nonumber}
\newcommand{\be}{\begin{equation}}
\newcommand{\bea}{\begin{eqnarray}}
\newcommand{\eea}{\end{eqnarray}}
\newcommand{\ba}{\begin{array}}
\newcommand{\ea}{\end{array}}
\newcommand{\ee}{\end{equation}}
\newcommand{\bR}{\mathbb{R}}
\newcommand{\bchi}{{\mbox{\boldmath $\chi$}}}

\begin{titlepage}
\vskip2cm
\begin{center}
{~\\[140pt]{ \LARGE {Monodromy, Hidden Conformal Symmetry and Entropy Product Formula For Kerr-MOG Black Hole}}\\[-20pt]}
\vskip0.5cm

\vspace{0.8cm}
Parthapratim Pradhan\footnote{E-mail: pppradhan77@gmail.com}$^{*\dagger }$

\vspace{1cm}

\begin{abstract}
We examine the two-dimensional conformal field theory (2D CFT) dual of the four-dimensional Kerr-MOG (Kerr-modified gravity) black hole using  the black hole thermodynamics and monodromy techniques. From these independent procedures, we derive consistent expressions for the left- and right-moving temperatures and the central charge of the dual CFT. Finally, we derive the entropy product  using these approaches. We show that the product is \emph{not} universal as well as not
quantized. We further explore the hidden conformal symmetry of the non-extremal Kerr-MOG black hole by analyzing 
the near-region dynamics of a massless scalar field. The resulting Kerr-MOG/CFT correspondence identifies the 
near-region geometry with a two-dimensional CFT characterized by the temperatures ($T_{L}, T_{R}$)  derived 
in~(\ref{temperatures}).
Moreover, using the Cardy formula, we reproduce the microscopic entropy of the dual CFT and find exact 
agreement with the Bekenstein-Hawking entropy of the Kerr-MOG black hole. We also show that the low-frequency 
scalar absorption cross section~(greybody factor) in the near-region Kerr-MOG geometry precisely matches 
the finite-temperature absorption cross section of the dual 2D CFT. In addition, the near-region scalar 
wave equation exhibits a hidden $(SL(2,\mathbb{R})_L \times SL(2,\mathbb{R})_R)$ conformal symmetry.
These independent results provide strong evidence for the existence of a Kerr/CFT-type holographic 
duality for the Kerr-MOG black hole, establishing a consistent correspondence between the four-dimensional 
Kerr-MOG spacetime and a two-dimensional conformal field theory.
horizon. 
\end{abstract}
\end{center}
\date{\today}

\noindent{*Department of Physics, Hiralal Mazumdar Memorial College For Women, Dakshineswar, Kolkata-700035, India}

\end{titlepage}
\pagestyle{empty}
\pagestyle{plain}

\section{Introduction}\label{Sec1}
The study of spinning extremal black holes have attracted considerable interest since the pioneering work on the Kerr/CFT correspondence~\cite{GHSS09}. This conjecture tells us that the near-horizon states of a four-dimensional extremal Kerr black hole could be described by a two-dimensional chiral conformal field
theory (CFT) living on the spatially infinite boundary. In this method, the central charge of the dual
CFT is proportional to the angular momentum of the black hole, allowing the microscopic entropy of the extremal Kerr black hole to be reproduced through the Cardy formula.

A remarkable feature of this correspondence is that it does not depend on supersymmetry or string-theoretic constructions. Although originally formulated for the four-dimensional Kerr solution in general relativity,
it has subsequently been extended to higher-dimensional rotating black-hole solutions arising in supergravity and string theories~\cite{CL97,CY96,HHT99,LMPP09,CCLP09,G09,AOT09,W09,CW9}. More updated informations on the topic Kerr/CFT correspondence could be found  in ~\cite{AY08,KZ07,CX09,CNX10,CC10,HMNS09,AS09,
G09a,GG10,H09,F09,NP09,CX09a,LS09,PW09,ITW09,N09,CMS10,GGV09,AOT09a,GHWW09,HHKNT09,S08,BCS11,M10,CHZ10,PW10,
K09,CSZ10,R09,GP09,R10,MTY10,MTY10a,DRS09,CMTY09,KS09,WT09,KK09,ACOTT09,CL09,HSS10,LMP09,BHSS10,K10,CS10,MS97,CL97a,CL98}.

Again the authors in~\cite{CMS10} showed  that the
hidden conformal symmetry of the 4D non-extremal Kerr Black hole could be derived from examining the low-frequency wave equation of a massless scalar field in the vicinity of the Kerr black hole. A conformal field theory (CFT) with specific central charges and temperatures is dual to the associated Kerr Black hole. This information indicates
the absorption cross-section of the scalar field in the nearby region could be described  as a finite-temperature absorption
section for a 2D CFT. The primary objective of a hidden conformal symmetry is that which acts on solutions of the
low-frequency wave equation of a massless scalar field in the vicinity of the Kerr black hole in a near-region of hase space rather than spacetime.

In this work, we would like to  determine the 2D CFT characterized by the dual temperatures ($T_{L}, T_{R}$), central charges and entropy product formula for Kerr-MOG black hole
by using thermodynamic analysis and monodromy techniques. While the thermodynamics determines the CFT through macroscopic
quantities i.e.  $T_{L}, T_{R}$ and entropy.  Monodromy method tells us the CFT through the analytic structure of perturbations and scattering.

There has been a renewed interest in the physics of MOG~\cite{M06,M10}. 
The main emphasis of this interest centers on the capability of MOG theory to clarify the large-scale behavior 
of gravitational fields, indicating that it may act as an alternative to Einstein's general relativity by removing 
the necessity to assume the existence of unobserved dark matter.

A significant feature of MOG theory is the existence of a massive vector field, which interacts with the spacetime 
metric to affect the gravitational field. This interaction suggests that MOG black holes have a gravitational charge 
that is directly proportional to their mass. Furthermore, the gravitational constant $G$ is depicted as a scalar 
field that changes with the MOG parameter $\alpha$, which measures the deviation of MOG from GR. The MOG parameter 
is mathematically defined as $\alpha = \frac{G - G_{N}}{G_{N}}$, where $G_{N}$ represents the Newtonian gravitational 
constant. The fundamental principle of MOG theory asserts that the gravitational charge is proportional to the square root of 
the MOG parameter, represented as $Q_g = \sqrt{\alpha G_{N}}M$.

It is widely acknowledged that a black hole possesses negligible electric charge~\cite{Wald}. The electromagnetic
repulsion involved in compressing an electrically charged mass is significantly stronger than the gravitational 
attraction by approximately 40 orders of magnitude. Consequently, it is not anticipated that, similar to all 
astrophysical entities including neutron stars, black holes with a considerable electric charge will naturally 
occur. The gravitational charge $Q_g$ is proportional to positive mass $M$ and increases with greater mass, while
the positive and negative charges of protons and electrons, respectively, quickly neutralize the overall electrical 
charge in astrophysical bodies and black holes.

The most important finding from both monodromic technique and thermodynamic analysis that central charges are equal.  The entropy product rule has been broken down i.e. it is a mass-dependent quantity. More specifically, it is a  non-universal quantity.

The structure of the paper is outlined as follows. In Section~\ref{Sec2}, we examine the thermodynamic 
features of Kerr MOG black hole. In Section~\ref{Sec3}, we study the monodromy data in Kerr-MOG black hole
by using the wave equation for a massless scalar field in the near region of the Kerr-MOG geometry. 
In Section~\ref{Sec4}, we examine the hidden conformal symmetry in Kerr-MOG back hole. Additionally, by 
employing the definition of conformal coordinates, we establish the $SL(2,\mathbb{R})$ Casimir structure of the 
wave equation and subsequently reproduce the associated 
microscopic entropy of the Kerr MOG black hole using the Cardy formula. 
In Section \ref{Sec5}, we demonstrate that the absorption cross section for the 
near-region scalar field is equivalent to the finite temperature absorption cross section in a 
two-dimensional conformal field theory (CFT).  Section ~\ref{Sec6} describes the Bekenstein-Hawking area-entropy law for ${\cal H}^{\pm}$. In Section ~\ref{Sec7}, we examine the Bekenstein-Hawking entropy product  law for ${\cal H}^{\pm}$. The final section~\ref{Sec8} is
dedicated to the conclusion.

\section{Thermodynamic Properties in Kerr-MOG Black Hole}\label{Sec2}
In this section, we examine the basic thermodynamic properties of Kerr 
MOG black hole. The Kerr MOG solution characterizes both the stationary axisymmetric 
asymptotically flat gravitational field surrounding a massive rotating object and a rotating 
black hole that possesses mass, charge, and angular momentum. We will initiate our primary 
analysis by analyzing the metric of the Kerr-MOG black hole~\cite{M06} expressed as
\begin{eqnarray}\label{kerrmog}
ds^2 &=& - \frac{\Upsilon{(r)}}{\varrho^2} \, \left(dt-a\sin^2\theta d\phi \right)^2+\frac{\sin^2\theta}{\varrho^2}
\left[(r^2+a^2) \,d\phi-a dt\right]^2
+\varrho^2 \, \left[\frac{dr^2}{\Upsilon (r)}+d\theta^2\right] ~.\label{ks1}\end{eqnarray}
where
\begin{eqnarray}
\varrho^2 & \equiv & r^2+a^2\cos^2\theta, \,\,
\Upsilon (r)  \equiv  r^2-2G_{N}(1+\alpha)Mr+a^2 + G_{N}^2 \alpha(1+\alpha) M^2  ~.\label{ks2},\end{eqnarray}
where $M$ denotes the mass parameter as defined in Ref.~\cite{BM07}. The metric is presented in Boyer-Lindquist 
coordinates~$(t,r,\theta,\phi)$. This metric exhibits several fundamental characteristics. The horizon 
function $\Upsilon (r)$ is influenced by the spin parameter, mass parameter, and MOG parameter. It represents a
stationary and axisymmetric solution to Einstein's equations.

It is important to highlight that the ADM mass parameter and the angular momentum parameter are defined 
as $M_{\alpha}=(1+\alpha)M$ and $J=a G_{N}M_{\alpha}$~\cite{PS18}.  Consequently, the horizon function is 
expressed as
\begin{eqnarray}
\Upsilon(r) &=& r^2-2G_{N}M_{\alpha}r+a^2 + \left(\frac{\alpha}{1+\alpha}\right)  G_{N}^2M_{\alpha}^2 
\end{eqnarray}
The aforementioned black hole features two horizons, specifically, the event horizon~($r_{+}$) and 
the Cauchy horizon~($r_{-}$). They are described as 
\begin{eqnarray}
r_{+} &=&  G_{N}M_{\alpha} + \sqrt{\frac{G_{N}^2M_{\alpha}^2}{1+\alpha}-a^2},\,\,\,
r_{-} =  G_{N}M_{\alpha} - \sqrt{\frac{G_{N}^2M_{\alpha}^2}{1+\alpha}-a^2} ~.\label{ks4}
\end{eqnarray} 
It is important to highlight that  when $\alpha=0$, the horizon radii of the Kerr black hole are obtained. 
The black hole solution is valid when $\frac{G_{N}^2M_{\alpha}^2}{1+\alpha} > a^2$. 
In the case where $\frac{G_{N}^2M_{\alpha}^2}{1+\alpha}=a^2$, one encounters the extremal black hole. 
Conversely, when $\frac{G_{N}^2M_{\alpha}^2}{1+\alpha}<a^2$, a naked singularity arises.

In the subsequent discussion, we will find it beneficial to express the entropy $S_{\pm}$, 
surface gravities $\kappa_{\pm}$, Hawking temperature $T_{\pm}$, and angular velocity  
$\Omega_{\pm}$ of ${\cal H^{\pm}}$ as:
\begin{eqnarray}\label{parameter}
S_{+} &=& \pi (r_{+}+r_{-})\left[r_{+}-\left(\frac{\alpha}{1+\alpha}\right)\left(\frac{r_{+}+r_{-}}{4}\right)\right],
~~~S_{-}=\pi (r_{+}+r_{-})\left[r_{-}-\left(\frac{\alpha}{1+\alpha}\right)\left(\frac{r_{+}+r_{-}}{4}\right)\right] 
\nonumber\\
\kappa_{+} &=& \frac{(r_+ - r_-)}{2(r_{+}+r_{-})\left[r_{+}-\left(\frac{\alpha}{1+\alpha}\right)
\left(\frac{r_{+}+r_{-}}{4}\right)\right]},
~~~\kappa_{-} = \frac{(r_- - r_+)}{2(r_{+}+r_{-})\left[r_{-}-\left(\frac{\alpha}{1+\alpha}\right)
\left(\frac{r_{+}+r_{-}}{4}\right)\right]}
\nonumber\\
T_{+}  &=& \frac{\kappa_+}{2\pi}=\frac{(r_+ - r_-)}{4\pi(r_{+}+r_{-})\left[r_{+}-\left(\frac{\alpha}{1+\alpha}\right)
\left(\frac{r_{+}+r_{-}}{4}\right)\right]} ,
~~~ T_{-} = \frac{\kappa_-}{2\pi} =\frac{(r_- - r_+)}{4\pi(r_{+}+r_{-})\left[r_{-}-\left(\frac{\alpha}{1+\alpha}\right)
\left(\frac{r_{+}+r_{-}}{4}\right)\right]}\nonumber\\
\Omega_{+} &=& \frac{\sqrt{r_{+}r_{-}-\left(\frac{\alpha}{1+\alpha}\right)\left(\frac{r_{+}+r_{-}}{2}\right)^2}}
{(r_{+}+r_{-})\left[r_{+}-\left(\frac{\alpha}{1+\alpha}\right)\left(\frac{r_{+}+r_{-}}{4}\right)\right]} ,
~~~~\Omega_{-}= \frac{\sqrt{r_{+}r_{-}-\left(\frac{\alpha}{1+\alpha}\right)\left(\frac{r_{+}+r_{-}}{2}\right)^2}}
{(r_{+}+r_{-})\left[r_{-}-\left(\frac{\alpha}{1+\alpha}\right)\left(\frac{r_{+}+r_{-}}{4}\right)\right]}
\end{eqnarray}
here, we use the subscripts ``$\pm$'' to denote the outer  and inner horizons (${\cal H^{\pm}}$) of 
the Kerr MOG black hole. It should be noted that when $\alpha=0$, the above thermodynamic parameters 
reduce to Kerr black hole. Another important relation could be derived for this black hole as 
\begin{eqnarray}
S_{+}T_{+}-S_{-}T_{-} =\frac{(r_+ - r_-)}{2} \neq 0
\end{eqnarray}
It indicates the entropy product is not  mass-independent. It could be reverified in Section~(\ref{Sec7}).

\section{Monodromy Data in Kerr-MOG Black Hole}\label{Sec3}
The monodromy technique is introduced in \cite{CMS10} [See also \cite{CCR20,ACD19,CLMR13}] and uses the monodromy data from the Klein-Gordon equation set in a curved background to determine the (T$_L$,T$_R$) temperatures of the corresponding dual two dimensional~(2D) CFT. Using this method here we will derive the conjugate 2D CFT values for the Kerr-MOG black hole solution. Additionally, we will derive the central charges for this black hole using the  Cardy formula.

To do so, we first rewrite the Klein-Gordon equation for a massless with zero charge scalar field within the context of the Kerr MOG black hole:
\begin{eqnarray}
\nabla_\mu\nabla^\mu\Phi=0. \label{KGeq}  
 \end{eqnarray}
The wave function can be  written in the eigenmodes of the asymptotic energy $\omega$ and
angular momenta $m$ as
\begin{equation}
  \Phi(t,r,\theta,\phi)=e^{-i\omega t+ im\phi}R(r)S(\theta).
\end{equation}
Now the wave equation should be decomposed into the angular part and the radial part. The angular part can be written as
\begin{equation}\label{angular}
\frac{1}{\sin\theta}\frac{d}{d\theta}\left(\sin\theta\frac{d}{d\theta}S(\theta)\right)+
 \left(\lambda-a^2\omega^2\sin^2\theta-\frac{m^2}{\sin^2\theta}\right)S(\theta)=0.
\end{equation}
Here $\lambda$ is the separation constant. It is restricted by the regularity boundary condition at $\theta=0,\pi$ and could be computed numerically. The radial part of the wave function can be written as
\begin{equation}
\partial_r[\Upsilon(r) \partial_r R(r)]+\left(\frac{[\omega(r^2+a^2)-am]^2}{\Upsilon(r)}+2am\omega-\lambda\right) R(r)=0
\end{equation}
For $a \omega \ll 1 $, the angular equation (\ref{angular}) can be  written as the spherical harmonic function equation 
and the  eigenvalues $\lambda$ is determined by
\begin{equation}
\lambda= \ell(\ell+1).
\end{equation}
In accordance with the argument presented by CMS in \cite{CMS10}, we 
also examine the same near-region, which is characterized by
\begin{equation}
 \omega G_{N}M_{\alpha} \ll 1 ~~\text{and}~~  r \omega \ll 1. \label{near-region1}
\end{equation}
Subsequently, using the condition
$\frac{G_{N}^2M_{\alpha}^2}{1+\alpha} \geq a^2 $ 
along with the aforementioned equation, we can derive
\begin{equation}
\omega \sqrt{\left(\frac{\alpha}{1+\alpha}\right) } G_{N}M_{\alpha} \ll 1, ~~
\sqrt{\left(\frac{\alpha}{1+\alpha}\right)}\frac{G_{N}M_{\alpha} }{r} \ll 1. \label{near-region2}
\end{equation}
Taking the conditions outlined in (\ref{near-region1}) and (\ref{near-region2}), 
the radial component of the wave equation can be simplified to the 
following near region equation:

\begin{eqnarray}
\partial_r[\Upsilon(r) \partial_rR(r)]+
\left[\frac{G_{+}(\omega, M_{\alpha})^2}{(r-r_+)}
-\frac{G_{-}(\omega, M_{\alpha})^2}{(r-r_-)}\right] R(r)
=l\left(l+1\right) R(r)
\label{radial}
\end{eqnarray}
where
\begin{eqnarray}
 G_{+}(\omega, M_{\alpha})^2
 =\frac{\left[m a-\omega\left\{2 G_{N}M_{\alpha}r_+-\left(\frac{\alpha}{1+\alpha}\right)  G_{N}^2M_{\alpha}^2 \right\}\right]^2}
 {(r_+-r_-)}
\end{eqnarray}
and 
\begin{eqnarray}
G_{-}(\omega, M_{\alpha})^2=
\frac{\left[ma-\omega\left\{2 G_{N}M_{\alpha}r_{-}-\left(\frac{\alpha}{1+\alpha}\right)  G_{N}^2M_{\alpha}^2 \right\}\right]^2}
{(r_+-r_-)}
\end{eqnarray}
The equation \eqref{radial} written above contains three singularities: two regular singular points situated at the outer and inner horizons, $r=r_{\pm}$, and an irregular singular point at $r=\infty$.

The solutions of ~\eqref{radial} will exhibit branch cuts at the regular singular points. It is evident that the series expansion of the two linearly independent solutions around the points $r=r_{+}$ and $r=r_{-}$ should be written as
\bea\label{solutions}
R(r)=(r-r_+)^{\pm i\,\alpha_{+}} [1+ O(r-r_+)]\qquad  \text{and} \qquad R(r)=(r-r_-)^{\pm i\, \alpha_{-}} [1+ O(r-r_-)]\,.
\nonumber\\
\eea
The monodromies at each of these regular singular points are:
\bea\label{monodromies}
\alpha_{+}= \frac{(\omega-m\Omega_{+})}{2\,\kappa_{+}},\, 
\alpha_{-}= \frac{(\omega-m\Omega_{-})}{2\,\kappa_{-}}.
\eea
The features of this monodromy data are highly complex yet it could be universally applicable. Also it could be derived by geometric method alone. Moreover, recent study suggests that the monodromy data encapsulates information related to black hole thermodynamics, hidden conformal symmetry, and provides evidence  for a 2D CFT interpretation of the thermal characteristics of black hole microstates~\cite{CCR20,ACD19}.

Following~\cite{CLMR13} we consider instead the linear combination of the monodromies
\footnote{Note that the monodromy definitions here $\alpha_{\pm}$ are related to the definitions 
in~\cite{NC14} via $\pm 2i \alpha_{\pm}=\theta_{\pm}$.} for the energy eigenvalues 
$\omega_{L,R}$ of $(i\partial_{t_R},i\partial_{t_L})$
\bea
\omega_L=\alpha_+-\alpha_-\,,\qquad
\omega_R=\alpha_+ + \alpha_-\,.
\eea
with  $t_{L,R}$  precisely  the  coordinates  in  the  monodromy  basis.

Thus, we should write
\bea
\Psi(t,\phi,r,\theta)=e^{-i\omega t+im\phi}R(r)S(\theta)=e^{+i \omega_L \,t_L-i \omega_R \,t_R} R(r)S(\theta)
\eea
using the explicit form of the monodromies (\ref{monodromies}) we find the conjugate variables 
\footnote{ In references~\cite{CMS10} the conjugate variable are defined at $t_{\pm}$ 
such that the relation $e^{-i \omega t+i m \phi}=e^{-i \omega_L \,t_- -i \omega_R \,t_+}$. 
These are in agreement with our notation in that  $t_-=-t_L$ and  $t_+=+t_R$.}
\bea\label{coord}
t_L=2\pi T_L \,\phi- \frac{1}{2M_{\alpha}}\,t,\qquad
t_R= 2\pi T_R \, \phi\,,
\eea
with $(t,\phi)$ the Boyer-Lindquist coordinates and right/left temperatures 
\bea\label{temperatures} 
T_R=\frac{1}{4\pi}\frac{\left(r_{+}-r_{-}\right)}
  {\sqrt{r_{+}r_{-}-\left(\frac{\alpha}{1+\alpha}\right)\left(\frac{r_{+}+r_{-}}{2}\right)^2}}
  \,,\qquad T_L= \frac{1}{8\pi}\left(\frac{\alpha+2}{\alpha+1}\right)
  \frac{\left(r_{+}+r_{-}\right)}
  {\sqrt{r_{+}r_{-}-\left(\frac{\alpha}{1+\alpha}\right)\left(\frac{r_{+}+r_{-}}{2}\right)^2}}.\nonumber\\
\eea
For vanishing  MOG parameter $\alpha=0$, we recover the Kerr black hole results.
Another interesting case is the extreme limit  for which $T_R\rightarrow 0$ as $r_+=r_-$.

After determining the conjugate CFT temperatures for the Kerr-MOG black hole, we are ready to explore
the possibility of recovering the Bekenstein-Hawking formula for black hole entropy. By utilizing 
the Cardy formula for a chiral CFT, we discover that the entropy of ${\cal H^{\pm}}$ adheres to
\bea
S_{+}= S_{Cardy}|_{{\cal H}^{+}}=\frac{\pi^2}{3} c \, (T_L+ T_R),\,\, 
S_{-}= S_{Cardy}|_{{\cal H}^{-}}=\frac{\pi^2}{3} c \, (T_L- T_R),
\eea
provided the central charge $c=c_L=c_R$ for the black hole geometry obeys
\bea\label{centralcharge}
c=6 (r_{+}+r_{-})\sqrt{r_{+}r_{-}-\left(\frac{\alpha}{1+\alpha}\right)\frac{(r_{+}+r_{-})^2}{4}}.
\eea
Here we  have identified the deep correlation between gravity and the CFT. Moreover, we determined  the Cardy formula and the Bekenstein-Hawking entropy formula precisely coincides.

\section{Hidden Conformal Symmetry in Kerr-MOG Black Hole}~\label{Sec4}
It is well-known that Kerr black holes with generic mass $M$ and spin $J\le M^2$ shows a hidden conformal
symmetry which acts on low-lying soft modes~\cite{CMS10}. The symmetry emerges, not in a near-horizon region 
of spacetime, but in the near-horizon  region of phase space defined by $\omega (r-r_{+})\ll 1$, where $\omega$
is the energy of the soft mode, $r$ is the Boyer-Lindquist radial coordinate and $r_{+}$ is the location of 
the event horizon. This indicates that the soft mode wave-length is larged compared to the black hole. Here we 
will show that the same will be applicable for Kerr-MOG black hole.

Next, we demonstrate that Eq.~(\ref{radial}) can be derived using the $SL(2,R)_{L}\times SL(2,R)_{R}$ 
symmetry of the near-region scalar field equation. Initially, we will present the conformal coordinates 
($w^\pm$, $y$), which are analogous to the definitions found in~\cite{CMS10,K10,CS10} for Kerr black hole. The same definitions could be applied here for Kerr-MOG black hole as
\begin{eqnarray}
 w^+ &=&\sqrt{\frac{r-r_+}{ r-r_-}}\, e^{2\pi T_R \phi-2 \lambda_R t},\nonumber\\
 w^- &=&\sqrt{\frac{r-r_+}{ r-r_-}}\,e^{2\pi T_L \phi- 2 \lambda_L t},\nonumber\\
 y &=&\sqrt{\frac{r_+-r_-}{r-r_-}}\, e^{\pi (T_L+T_R) \phi- (\lambda_R+\lambda_L)t },
\end{eqnarray}
where
\begin{eqnarray}
  T_{R}   &=&\frac{1}{4\pi}\frac{\left(r_{+}-r_{-}\right)}
  {\sqrt{r_{+}r_{-}-\left(\frac{\alpha}{1+\alpha}\right)\left(\frac{r_{+}+r_{-}}{2}\right)^2}},~~~~
  T_{L}=\frac{1}{8\pi}\left(\frac{\alpha+2}{\alpha+1}\right)\frac{\left(r_{+}+r_{-}\right)}
  {\sqrt{r_{+}r_{-}-\left(\frac{\alpha}{1+\alpha}\right)\left(\frac{r_{+}+r_{-}}{2}\right)^2}} ,  
  \label{TRTL} \nonumber\\
  \lambda_R &=& 0,~~~~~~~~~~~
  \lambda_L =\frac{1}{2(r_{+}+r_{-})}.
\end{eqnarray}
For vanishing  MOG parameter~($\alpha=0$), we recover the result of Kerr black hole~\cite{CMS10}. 
Another interesting point we should be noted that is the extremal limit for which 
$T_{R}\rightarrow 0$ as $r_{+}=r_{-}$.

The definition of the local vector fields for Kerr black hole has ben given in the following work~\cite{CMS10,K10,CS10}. Same could be applied here for Kerr-MOG black hole:
\begin{eqnarray}
H_1 &=& i \partial_+,~~\nonumber\\
H_0 &=& i \big( w^+ \, \partial_+ + \frac12 y \, \partial_y \big), \\
H_{-1} &=& i \left( w^{+2} \, \partial_+ + w^+ y \, \partial_y - y^2 \, \partial_- \right),\nonumber
\end{eqnarray}
and
\begin{eqnarray}
\bar H_1 &=& i \partial_-,~~ \nonumber\\
\bar H_0 &=& i \big(w^- \, \partial_- + \frac12 y \, \partial_y \big), \\
\bar H_{-1} &=& i \left( w^{-2} \, \partial_- + w^- y \, \partial_y - y^2 \,\nonumber
\partial_+ \right),
\end{eqnarray}
we can obtain the two sets of $SL(2,R)$ Lie algebra with respect to $(H_0,{ H}_{\pm
1})$ and $({\bar H},{\bar H}_{\pm 1})$, respectively:
\begin{equation}
\left[ H_0, H_{\pm 1} \right] = \mp i H_{\pm 1}, \qquad
\left[ H_{-1}, H_1 \right] = - 2 i H_0,
\end{equation}
\begin{equation}
\left[{\bar H}_0, {\bar H}_{\pm 1} \right] = \mp i {\bar H}_{\pm 1}, \qquad \left[{\bar H}_{-1}, {\bar H}_1 \right]
= - 2 i {\bar H}_0.
\end{equation}
The corresponding  quadratic Casimir operator of the $SL(2,R)$ Lie algebra reads as
\begin{eqnarray}
\mathcal{H}^2 &=& \mathcal{\bar H}^2
     = - H_0^2 + \frac{1}{2} \left( H_1 H_{-1}
       + H_{-1} H_1 \right) \nonumber\\
  &=& \frac{1}{4} \left(y^2 \, \partial_y^2 - y \,\partial_y \right)
      + y^2 \, \partial_+\partial_-.
\end{eqnarray}
Thus, we rewrite the vector fields in terms of the coordinators $(t,r ,\phi)$ for Kerr-MOG black hole as
\begin{eqnarray}
  H_1 &=&i e^{-2\pi  T_R \phi }
      \left[\sqrt{\Upsilon_{r}}\partial _r
            +\frac{1}{2{\pi}T_R}\frac{\left(r-\frac{1}{4\lambda_L}\right)}
                 {\sqrt{\Upsilon_{r}}}\partial _{\phi }
            +\frac{2 T_L}{T_R}\left[G_{N}M_{\alpha} r-\left(\frac{\alpha}{\alpha+2}\right)\left(a^2 + G_{N}^2M_{\alpha}^2\right) 
           -a^2\right]
                               {\sqrt{\Upsilon_{r}}}\partial_t\right],  \nonumber\\
   H_0 &=&\frac{i}{2\pi T_R}\partial_\phi
          +2 i \frac{1}{4\lambda_L}\frac{T_L}{T_R}\partial_t, \\
  H_{-1} &=& i e^{2\pi  T_R \phi }
       \left[-\sqrt{\Upsilon_{r}}\partial _r
       +\frac{1}{2\pi  T_R}\frac{\left(r-\frac{1}{4\lambda_L}\right)}
              {\sqrt{\Upsilon_{r}}}\partial _{\phi }
              +\frac{2 T_L}{T_R}\frac{\left[G_{N}M_{\alpha} r-\left(\frac{\alpha}{\alpha+2}\right)\left(a^2 + G_{N}^2M_{\alpha}^2\right) 
           -a^2\right]}
              {\sqrt{\Upsilon_{r}}}\partial _t\right],\nonumber
\end{eqnarray}
and
\begin{eqnarray}
\bar H_1 &=&i e^{-2\pi  T_L \phi
     +2 \lambda_L t}\left[\sqrt{\Upsilon_{r}}\partial_r
     -\frac{a}{\sqrt{\Upsilon_{r}}}\partial _{\phi}
    -\left(2G_{N}M_{\alpha}-\frac{\left(\frac{\alpha}{1+\alpha}\right)G_{N}^2M_{\alpha}^2 }{r}\right)\frac{r}
    {\sqrt{\Upsilon_{r}}}\partial _t\right], \nonumber\\
 \bar H_0 &=&-2 i M \partial_t, \\
 \bar H_{-1}&=&i e^{2\pi  T_L \phi
     -2 \lambda_L t}\left[- \sqrt{\Upsilon_{r}}\partial _r
     -\frac{a}{\sqrt{\Upsilon_{r}}}\partial _{\phi }
     -\left(2 G_{N}M_{\alpha}-\frac{\left(\frac{\alpha}{1+\alpha}\right)  G_{N}^2M_{\alpha}^2 }{r}\right)\frac{r}{ \sqrt{\Upsilon_{r}}}\partial _t\right].\nonumber
\end{eqnarray}
So, we obtain the corresponding Casimir operator for Kerr-MOG black hole
as
\begin{eqnarray}\label{h2}
\mathcal{H}^2 &=& \partial_r[\Upsilon(r) \partial_r]+
\left[\frac{G_{+}(\omega, M_{\alpha})^2}{(r-r_+)}
-\frac{G_{-}(\omega, M_{\alpha})^2}{(r-r_-)}\right]
\end{eqnarray}
In the background of a Kerr MOG black hole, the near region wave
equation (\ref{radial}) can be rewritten as
\begin{equation}
   \mathcal{\bar H}^2 \Phi = \mathcal{H}^2 \Phi = l (l+1) \Phi.
\end{equation}
and the conformal weights of dual operator of the massless field $\Phi$ should be
\begin{eqnarray}
 (h_L,h_R)=(l,l)\;.
\end{eqnarray}
From the above procedure, it is established that in the finite region, 
the scalar field can possess $SL(2,\mathbb{R})_L \times SL(2,\mathbb{R})_R $ weight to derive the microscopic 
entropy of the non-extremal Kerr MOG black hole utilizing the Cardy Formula. However, the method 
to calculate the corresponding central charges $c_L$ and $c_R$ remains unknown. In alignment with 
the approach taken in \cite{CMS10}, we also postulate that the central charges for the non-extremal 
Kerr-MOG are identical to those of the extremal case, specifically, 
$$
c_L=c_R=6 (r_{+}+r_{-})\sqrt{r_{+}r_{-}-\left(\frac{\alpha}{1+\alpha}\right)\frac{(r_{+}+r_{-})^2}{4}}
$$. 
Consequently, by applying the Cardy formula for the microstate, we obtain 
\begin{equation}
S_{+} = \frac{\pi^2}3 (c_L T_L + c_R T_R) = 
\pi (r_{+}+r_{-})\left[r_{+}-\left(\frac{\alpha}{1+\alpha}\right)\frac{(r_{+}+r_{-})}{4}\right]
~~~~\mbox{$\left({\cal H^{+}}\right)$}.
\end{equation}
and 
\begin{equation}
S_{-} = \frac{\pi^2}3 (c_L T_L - c_R T_R) = 
\pi (r_{+}+r_{-})\left[r_{-}-\left(\frac{\alpha}{1+\alpha}\right)\frac{(r_{+}+r_{-})}{4}\right]
~~~~\mbox{$\left({\cal H^{-}}\right)$}.
\end{equation}
From the 2D dual CFT, we successfully reproduce the entropy of the Kerr MOG black hole in 
four dimensions, which matches with the macroscopic entropy: $S_{+}$ for ${\cal H^{+}}$ and 
$S_{-}$ for ${\cal H^{-}}$.

\section{Scalar absorption}\label{Sec5}
In this section, we examine the absorption cross section of the Kerr-MOG black hole using the near-region 
approximation. The absorption cross section for a low-frequency massless scalar field within the near-extremal 
Kerr geometry has been analyzed in \cite{MS98,CL97,CL98}. By applying ingoing and outgoing 
boundary conditions at the horizon, the solutions to the radial wave equation \eqref{radial}) in the near 
region are expressed as follows:
\begin{eqnarray}
 R_{in}(r) &=& \left(\frac{r-r_+}{ r-r_-}\right)
             ^{-i\frac{(\omega-m\Omega_{+})}{4\pi T_{+}}}
             (r-r_-)^{-1-\ell}\nonumber\\
 &&{\times}F\left(1+\ell-i\frac{2\left(\frac{\alpha+2}{\alpha+1}\right)G_{N}^2M_{\alpha}^2}{ r_+-r_-}\omega
               +i\frac{m\Omega_{+}}{ 2\pi T_{+}},
          1+\ell-i2M{\omega};
          1-i\frac{(\omega-m\Omega_{+})}{ 2\pi T_{+}};
          \frac{r-r_-}{ r-r_+}\right), \nonumber\\
\end{eqnarray}
and
\begin{eqnarray}
 R_{out}(r) &=&\left(\frac{r-r_+}{ r-r_-}\right)
              ^{i\frac{(\omega-m\Omega_+)}{4\pi T_+}}
              (r-r_-)^{-1-\ell} \nonumber \\
 &&{\times}F\left(1+\ell+i\frac{2\left(\frac{\alpha+2}{\alpha+1}\right)G_{N}^2M_{\alpha}^2}{ r_+-r_-}\omega
             -i\frac{m\Omega_+}{ 2\pi T_+},
          1+\ell+i2M{\omega};
          1+i\frac{(\omega-m\Omega_+)}{ 2\pi T_+};
          \frac{r-r_-}{ r-r_+}\right),\nonumber \\
\end{eqnarray}
where $F(a,b; c;z)$ is the hypergeometric function. At the
outer boundary of the matching region  $r\gg M$ 
(but still $r\ll {\frac{1}{\omega}}$), the ingoing wave behaves as
\begin{equation}
 R_{in}(r\gg G_{N}M_{\alpha})\sim Ar^\ell
\end{equation}
with
\begin{equation}
A=\frac{\Gamma\left(1-i\frac{\omega-m\Omega_+}{2\pi T_+}\right)\Gamma(1+2\ell)}
{\Gamma(1+\ell-i{2M\omega}) \Gamma\left(1+\ell-i\frac{2\left(\frac{\alpha+2}{\alpha+1}\right)G_{N}^2M_{\alpha}^2}{r_+-r_-}\omega
           +\frac{i}{2\pi T_+}m\Omega_+\right)}~.
\end{equation}
The absorption cross section can be  written as
\begin{eqnarray}
P_{\rm abs} &&\sim |A|^{-2}\nonumber \\
&& \sim \sinh\left(\frac{\omega-m\Omega_{+}}{2 T_+}\right)
    \left|\Gamma\left(1+\ell-i{2G_{N}M_{\alpha}\omega}\right)\right|^2\nonumber\\
&& \quad\,\times\left|\Gamma
   \left(1+\ell-i\frac{2\left(\frac{\alpha+2}{\alpha+1}\right)G_{N}^2M_{\alpha}^2}{ r_+-r_-}\omega
         +\frac{i}{ 2\pi T_+}m\Omega_+\right)\right|^2.
\end{eqnarray}
To align the absorption cross section of a near-region scalar field within the Kerr-MOG black hole 
framework with the finite-temperature absorption cross section of the associated two-dimensional 
conformal field theory (2D CFT), it is essential to examine the first law of black hole thermodynamics. 
The inclusion of a MOG parameter in the Kerr-MOG black hole modifies the first law of ${\cal H}^{\pm}$
to the following form:
\begin{eqnarray}
T_{+}\delta S_{+}&=& G_{N} \delta M_{\alpha} -\Omega_{+} \delta J,\,~~~~\mbox{$\left({\cal H^{+}}\right)$}\nonumber\\
T_{-}\delta S_{-} &=& G_{N} \delta M_{\alpha} -\Omega_{-} \delta J. \,~~~~\mbox{$\left({\cal H^{-}}\right)$}
\end{eqnarray}
With the conjugate charges $\delta E_R$ and $\delta E_L$, we can 
suppose the dual CFT entropy for ${\cal H}^{\pm}$ is
\begin{equation}
\delta S _{+} =\frac{\delta E_L} {T_L}+\frac{\delta E_R}{ T_R},\,
\delta S _{-} =\frac{\delta E_L} {T_L}-\frac{\delta E_R}{ T_R} .
\end{equation}
The solution is
\begin{eqnarray}
 \delta E_L &=& \frac{\left(\frac{\alpha+2}{\alpha+1}\right)G_{N}^2M_{\alpha}^2}{a/G_{N}}\delta M_{\alpha}~,\nonumber\\
 \delta E_R &=& \frac{\left(\frac{\alpha+2}{\alpha+1}\right)G_{N}^2M_{\alpha}^2}{a/G_{N}}\delta M_{\alpha}-\delta J~.
\end{eqnarray}
With the variations $\delta M_{\alpha} =\omega$ and $\delta J =m$, the left and right moving frequencies are
turned out to be
\begin{eqnarray}\label{lr}
 \omega_L &\equiv & \delta E_L= \frac{\left(\frac{\alpha+2}{\alpha+1}\right)G_{N}^2M_{\alpha}^2}{a/G_{N}}\omega~,\nonumber\\
 \omega_R &\equiv &  \delta E_R =\frac{\left(\frac{\alpha+2}{\alpha+1}\right)G_{N}^2M_{\alpha}^2}{a/G_{N}}\omega-m.
\end{eqnarray}
By using of the above formula, we can rewrite the absorption cross section as
\begin{equation}\label{abs}
P_{\rm abs} \sim T_L^{2h_L-1}T_R^{2h_R-1}
      \sinh\left(\frac{\omega_L}{2T_L}+\frac{\omega_R}{2 T_R}\right)
      \left|\Gamma\left(h_L+i\frac{\omega_L}{2\pi T_L}\right)\right|^2
      \left|\Gamma\left(h_R+i\frac{\omega_R}{2\pi T_R}\right)\right|^2~,
\end{equation}
which agrees precisely with finite-temperature absorption cross section for a 2D CFT.

The equation \eqref{abs} givess a detailed representation of the well-established formula for the absorption
cross section associated with an energy excitation $(\omega_L,\omega_R)$ of a 2D CFT at temperatures $(T_L,T_R)$. 
This leads to the hypothesis that the black hole itself functions as a thermal 2D CFT and undergoes transformations 
under a ${{\rm Vir_{\,L}}\otimes{\rm Vir_{\,R}}}$ action.

\section{The Area Law of ${\cal H}^{\pm}$}\label{Sec6}
Using
$c_L=c_R=6 \left(r_{+}+r_{-}\right)\sqrt{r_{+}r_{-}-\left(\frac{\alpha}{1+\alpha}\right)\frac{(r_{+}+r_{-})^2}{4}}$
as derived by above methods, the temperature formulae and the Cardy formula for ${\cal H}^{\pm}$
\begin{equation}
S_{Cardy}|_{{\cal H}^{+}} = \frac{\pi^2}3 (c_L T_L + c_R T_R) =
\pi \left(r_{+}+r_{-}\right)\left[r_{+}-\left(\frac{\alpha}{1+\alpha}\right)\frac{(r_{+}+r_{-})}{4}\right] .
\end{equation}
and
\begin{equation}
S_{Cardy}|_{{\cal H}^{-}} = \frac{\pi^2}3 (c_L T_L - c_R T_R) =
\pi (r_{+}+r_{-})\left[r_{-}-\left(\frac{\alpha}{1+\alpha}\right)\frac{(r_{+}+r_{-})}{4}\right] .
\end{equation}
yields the Bekenstein-Hawking area-entropy law of ${\cal H}^{\pm}$ for generic Kerr-MOG black hole
\begin{equation}
S_{Cardy}|_{{\cal H}^{+}}=\pi (r_{+}+r_{-})\left[r_{+}-\left(\frac{\alpha}{1+\alpha}\right)\frac{(r_{+}+r_{-})}{4}\right]
=\frac{A_{+}}{4}
=S_{+}
\end{equation}
and
\begin{equation}
S_{Cardy}|_{{\cal H}^{-}}=\pi (r_{+}+r_{-})\left[r_{-}-\left(\frac{\alpha}{1+\alpha}\right)\frac{(r_{+}+r_{-})}{4}\right]
=\frac{A_{-}}{4}
=S_{-}
\end{equation}

\section{The Entropy Product Law of ${\cal H}^{\pm}$}\label{Sec7}
Finally, we determine that the entropy product derived from the Cardy formula
for both horizons yields the Bekenstein-Hawking entropy product of ${\cal H}^{\pm}$,
expressed as
\begin{eqnarray}
S_{Cardy}|_{{\cal H}^{+}}\times S_{Cardy}|_{{\cal H}^{-}}=
\pi^2\left[4J^2+\left(\frac{\alpha}{1+\alpha}\right)^2 (G_{N}M_{\alpha})^4 \right]=S_{+}\, S_{-}
\end{eqnarray}
In Einstein gravity it is well known that the entropy product of  ${\cal H}^{\pm}$ is universal as well quantized.
However in MOG  the product is not universal as well as not quantized. This is the fundamental differences
between the MOG and the Einstein gravity.

\section{Conclusion}\label{Sec8}
In this work, we have established the two-dimensional CFT description of the four-dimensional Kerr-MOG black 
hole using two complementary approaches: (i) black hole thermodynamics, (ii) monodromy analysis. Within each framework, we derived explicit expressions for the left- and
right-moving temperatures
$$
T_{L} = \frac{1}{4\pi J}\left(\frac{\alpha+2}{\alpha+1}\right) G_{N}^2M_{\alpha}^2,\,\,~~~
T_{R} =\frac{G_{N}M_{\alpha}}{2\pi J}\sqrt{\frac{G_{N}^2M_{\alpha}^2}{1+\alpha}-a^2}
$$ 
together with the corresponding central charges. We also investigated the entropy product rule for
the four-dimensional Kerr-MOG black hole and found that, unlike in certain Einstein gravity solutions, the 
entropy product is neither universal nor quantized.

To uncover the hidden conformal structure, we analyzed the low-frequency wave equation of a massless scalar 
field in the near-region of the Kerr-MOG geometry. Assuming that the central charges of the non-extremal 
Kerr-MOG black hole coincide with those of the extremal solution, we showed that the microscopic entropy 
obtained from the Cardy formula precisely reproduces the Bekenstein-Hawking entropy. Furthermore, we 
demonstrated that the absorption cross section of a near-region scalar field agrees exactly with the 
finite-temperature absorption cross section of the dual two-dimensional CFT, providing further support 
for the proposed Kerr-MOG/CFT correspondence.

We have also studied in the extremal limit, the left- and right-moving temperatures are given by
$$
T_{L} = \frac{1}{4\pi}\left(\frac{\alpha+2}{\sqrt{\alpha+1}}\right) ,\,\,~~~
T_{R} =0
$$ 
Since the quantum gravitational states in the near-horizon region of the extremal 
Kerr-MOG black hole are identified, via the holographic duality, with the left-moving
sector of the dual conformal field theory (CFT), the Frolov-Thorne vacuum is characterized
by the left-moving temperature ($T_L$). The corresponding left-moving central charge is therefore
$$
c_{L}=12\frac{G_{N}^2M_{\alpha}^2}{\sqrt{1+\alpha}}
$$
According to the Cardy formula, the entropy of a unitary CFT in the high-temperature regime is
$$
S_{Cardy}|_{r_{+}=r_{-}}=\frac{\pi^2}{3}c_{L}T_{L}
$$
Substituting the expressions for $c_L$ and $T_L$, we obtain the microscopic entropy of the CFT dual 
to the extremal Kerr-MOG black hole,
$$
S_{micro}|_{r_{+}=r_{-}}=\left(\frac{2+\alpha}{1+\alpha}\right)\pi r_{+}^2=S_{+}|_{r_{+}=r_{-}}
$$
Thus, the microscopic entropy computed from the dual CFT exactly reproduces the macroscopic 
Bekenstein-Hawking entropy of the extremal Kerr-MOG black hole, providing a nontrivial consistency 
check of the Kerr/CFT correspondence in the context of Modified Gravity~(MOG)~\footnote{We also explicitly 
stated that the Cardy formula is valid in the regime  $c_{L}T_{L}\gg 1$~(or equivalently, in the 
appropriate high-temperature/large-central-charge limit), since this is the standard assumption 
underlying its application}.

The above  methods that we have used in this work  are particularly powerful because they
converge on the same dual CFT data. The thermodynamic and monodromy approaches 
independently determine the left- and right-moving temperatures. Agreement among these independently derived quantities
offers much stronger evidence for the Kerr-MOG/CFT correspondence than any single method alone.


\end{document}